\documentclass[aps,prd,showpacs,11pt,floatfix,nofootinbib,runinaddress]{revtex4-1}
\usepackage[utf8]{inputenc}               
\usepackage{amsmath}
\usepackage{mathtools}
\usepackage[T1]{fontenc}                  
\usepackage{amssymb}
\usepackage{graphicx}
\usepackage{hyperref}
\usepackage{braket}
\usepackage{dsfont}
\usepackage{siunitx}
\usepackage{mathtools, slashed}

\graphicspath{{Figs/}}

\begin{document}

\title{The Roper resonance in a finite volume}

\author{Daniel Severt$^1$}
\email{severt@hiskp.uni-bonn.de}

\author{Ulf-G. Mei{\ss}ner$^{1,2,3}$}
\email{meissner@itkp.uni-bonn.de}


\affiliation{
$^1$\mbox{Helmholtz-Institut f\"ur Strahlen- und
             Kernphysik and Bethe Center for Theoretical Physics,}\\
             \mbox{Universit\"at Bonn,  D-53115 Bonn, Germany}\\
$^2$\mbox{Institute for Advanced Simulation, Institut f\"ur Kernphysik
               and J\"ulich Center for Hadron Physics,}\\
           \mbox{Forschungszentrum J\"ulich, D-52425 J\"ulich, Germany}\\    
$^3$\mbox{Tbilisi State University, 0186 Tbilisi, Georgia}
}

\begin{abstract}\noindent 
  We calculate the energy levels corresponding to the Roper resonance based on
  a two-flavor chiral effective Lagrangian for pions, nucleons, deltas and the
  Roper resonance at leading one-loop order. We show that the Roper mass can be
  extracted from these levels for not too large lattice volumes.
\end{abstract}

\maketitle

\section{Introduction}
In the last years a lot of work has been done to understand the hadron spectrum as it emerges from Quantum Chromodynamics (QCD).
However, the excited baryon spectrum of QCD is still not very well understood and requires further theoretical investigations. 
At low energies chiral perturbation theory (ChPT) has proven to be an important tool to describe hadrons and their interactions,
especially in the Goldstone boson sector. The inclusion of baryon states in ChPT is also possible and baryon
chiral perturbation theory (BChPT) is widely and successfully used today. BChPT requires a more sophisticated
approach because of the breakdown of the power counting due to the inclusion of these heavy degrees of freedom and their
large masses. This issue can, however, be resolved either by using the so-called heavy-baryon approach or a suitably chosen
renormalization scheme within an explicitly Lorentz-invariant formalism, like the infrared regularization (IR) or
extended-on-mass shell (EOMS) approaches, see e.g. the review~\cite{Bernard:2007zu}.
This allows to investigate the properties of a few low-lying excited states. Unitarization methods allow to address
more meson and baryon resonances, however, at the cost of introducing some model-dependence as various unitarization
schemes can be employed. To access a larger part of the spectrum, a different approach is required. Lattice QCD is a first
principles method that allows to calculate the hadron spectrum
from the underlying fundamental quark and gluon fields. Calculations do an outstanding job in describing the
lowest-lying hadron states. With ever increasing computational power, improved algorithms and refined finite volume methods,
especially hadron ground states are simulated more and more precisely on the lattice. Excited states are more difficult to
access, though there has been some visible progress in the last years. The present state of the art is reviewed in
Ref.~\cite{Edwards}.

One excited state that is of perticular interest is the Roper resonance, which was found in 1964 using a partial wave analysis
of pion-nucleon scattering data \cite{Roper:1964zza}. It is a spin-$1/2$ state with positive parity (like the nucleon)
and with a mass of around $m_R = \SI{1.44}{\GeV}$\footnote{This is the less reliable Breit-Wigner mass \cite{Tanabashi:2018oca}.}
it lies slighty above the delta resonance. The most remarkable feature
of this low-lying baryon resonance are its  decays. Besides the decay into a pion and a nucleon, it  also decays
into a nucleon and two pions (via the $\Delta N$ and $N\sigma$ intermediate states)  with a branching fraction
comparable to the $N\pi$ mode. This three particle final state becomes important in
lattice simulations involving three or more hadrons, see Ref.~ \cite{Rusetsky:2019gyk} for a recent review.
It is also worth noting that there are going experimental programs to map out the electromagnetic structure
of the Roper resonance, in particular through electro-excitatation and related theoretical studies, see e.g.
Refs.~\cite{Bauer:2014cqa,Gelenava:2017mmk,Golli:2017nid,Wu:2017qve}.

A dedicated lattice QCD study of the Roper using both quark and hadron interpolators was performed in Ref.~\cite{Lang:2016hnn},
see also Ref.~\cite{Liu:2016uzk}. In Ref.~\cite{Lang:2016hnn}  a number
of three-quark interpolating fields was supplemented by  operators for $N\pi$ in P-wave and $N\sigma$
in S-wave. In the center-of-momentum frame  three eigenstates below 1.65~GeV were found.
No eigenstate corresponding to the Roper at $m_R = 1.44\,$GeV is found, which indicates that $N\pi$ elastic
scattering alone does not render a low-lying Roper. Coupling with other channels, most
notably with $N\pi\pi$, seems to be important for generating the Roper resonance. The study of the 
coupled-channel scattering  including a three-particle decay $N\pi\pi$ remains a challenge.  

Here, we follow another path. An effective field theory treatment of the Roper resonance has already been established.
In order to improve the investigation of the Roper on the lattice, a finite volume calculation of the system is performed.
The Roper is placed in a finite cubic box of size $L$ and we study the difference between its energy spectrum in the
infinite volume and finite volume case, i.e. the
finite volume corrections of the Roper resonance. Due to the presence of a narrow resonance, the energy levels in the box show a
very characteristic behaviour near the resonance energy. The energy levels get shifted when the box size $L$ is changed, but
they do not cross each other.  This is the so-called ``avoided level crossing''~\cite{Wiese:1988qy}.

In this work we want to find out if this behavior can also be seen in the energy levels of the Roper system. To do so, we
study the finite volume corrections of the self-energy of the Roper up to third chiral order $\mathcal{O}(p^3)$ and
perform a fit of the energy levels. A similar study has already been done for the delta resonance in \cite{Bernard:2007cm}
and we treat the Roper resonance accordingly.

Our manuscript is organized as follows: In Sec.~\ref{SecEffLagr}, we display the effective chiral two-flavour
Lagrangian of pions and baryons (nucleon, delta, Roper resonance) underlying our calculations. In Sec.~\ref{sec:ropercont},
we calculate the self-energy of the Roper resonance in the continuum volume.  The calculation of Roper self-energy in the finite
volume is given in Sec.~\ref{sec:finvol}. The results for the energy levels of the Roper and the pertinent discussion are
given in Sec.~\ref{sec:res}. We end with a short summary and outlook in Sec.~\ref{sec:out}. Some technicalities are relegated
to the appendices.

\section{Effective Lagrangian} \label{SecEffLagr}
First, we discuss the  chiral effective Lagrangian that we need for
our calculations. It is taken from Ref.~\cite{Gegelia:2016xcw} (for earlier related work,
see e.g. Refs.~\cite{Beane:2002ud,Borasoy:2006fk,Djukanovic:2009gt,Long:2011rt}) and is given by 
\begin{align}
\mathcal{L}_{\text{eff.}} = \mathcal{L}_{\pi \pi} + \mathcal{L}_{\pi N} + \mathcal{L}_{\pi R} + \mathcal{L}_{\pi \Delta}
+ \mathcal{L}_{\pi N \Delta} + \mathcal{L}_{\pi N R} + \mathcal{L}_{\pi \Delta R} \; .
\end{align}
The dynamical degrees of freedom are pions ($\pi$), nucleons ($N$), the delta ($\Delta$) and the Roper resonance ($R$).
We restrict ourselves to flavor $\text{SU}(2)$ and work in the isospin limit ($ m_u = m_d \equiv \hat{m} $).
In what follows, we work to leading one-loop order, $\mathcal{O}(p^3)$, where $p$ denotes a small momentum or mass.
We count the pion mass as well as the mass differences $m_R-m_N$, $m_\Delta-m_N$ and $m_R-m_\Delta$ as of order $p$.
When going to higher orders, this naive counting requires modification as detailed in Ref.~\cite{Gegelia:2016xcw}.
Now let us enumerate the contributions required for the $\mathcal{O}(p^3)$ calculation of the Roper self-energy. The relevant
terms from the mesonic Lagrangian are
\begin{align}
\mathcal{L}^{(2)}_{\pi \pi} &= \frac{F^2}{4} \text{Tr} \left( \partial_{\mu} U \partial^{\mu} U^{\dagger} \right)
+ \frac{F^2}{4} \text{Tr} \left( U \chi^{\dagger} + \chi U^{\dagger} \right) \; ,
\end{align}
where $U$ is a $2 \times 2$-matrix that contains the pion fields, $F$ is the pion decay constant in the chiral limit,
which will later be identfied with the physical pion decay constant $F_\pi$. Further, $\chi$ is the external scalar source
which is given by the diagonal matrix 
\begin{align}
	\chi = 
	\begin{pmatrix}
	M_{\pi}^2 & 0 \\
	0 & M_{\pi}^2 \\
	\end{pmatrix} \; ,
\end{align}
with the pion mass $M_{\pi}$ (we have already identified the leading term in the quark mass expansion of the
pion mass with its physical value). The leading order (LO) terms of chiral dimension one and one required next-to-leading
order (NLO) term of chiral dimension two containing pion fields and the spin-$1/2$ baryons read 
\begin{align}
\mathcal{L}^{(1)}_{\pi N} &= \bar{\Psi}_{N}^{\phantom{()}}  \left( i \slashed{D} - m_{N 0}^{\phantom{1}} + \frac{1}{2} g_{A}^{\phantom{()}} \slashed{u}
\gamma_5 \right) \Psi_{N}^{\phantom{()}}  \; , \notag \\
\mathcal{L}^{(1)}_{\pi R} &= \bar{\Psi}_{R}^{\phantom{()}}  \left( i \slashed{D} - m_{R 0}^{\phantom{1}} + \frac{1}{2} g_{R}^{\phantom{()}} \slashed{u}
\gamma_5 \right) \Psi_{R}^{\phantom{()}} \; , \notag\\ 	
\mathcal{L}^{(2)}_{\pi R} &= c_1^R \bar{\Psi}_{R}^{\phantom{()}} \text{Tr} \left(  \chi_+ \right) \Psi_{R}^{\phantom{()}} \; . \label{PiRNLagr}
\end{align}
Here, $\Psi_{N}$ and $\Psi_{R}$ are the isospin doublet fields with chiral limit masses $m_{N 0}$ and $m_{R 0}$ of the nucleon and the
Roper resonance, respectively. The interaction of these fields with the pion field is characterised by the axial couplings
$g_A$ and $g_R$ and the chiral vielbein 
\begin{align}
u_{\mu} = i \left( u^{\dagger} \partial_{\mu} u - u \partial_{\mu} u^{\dagger} \right) \; , 
\end{align}
where $ u = \sqrt{U} $. The last equation in (\ref{PiRNLagr}) denotes a term of the second order pion-Roper Lagrangian with the
low-energy constant (LEC) $c_1^R $ and $\chi_{+} = u^{\dagger} \chi u^{\dagger} + u \chi^{\dagger} u$. This term is required in
the calculation of the Roper self-energy to be discussed below. Since we are only interested in
strong interaction processes we leave out all other external sources. The covariant derivative is then given by 
\begin{align}
	D_{\mu} \Psi_{N/R} = \left( \partial_{\mu}  + \Gamma_{\mu} \right) \Psi_{N/R} \; , 
\end{align}
with 
\begin{align}
	\Gamma_{\mu} = \frac{1}{2} \left \lbrace u^{\dagger} \partial_{\mu} u + u \partial_{\mu} u^{\dagger} \right \rbrace \; .
\end{align}
The spin-$3/2$ delta resonances are introduced as usual in terms of Rarita-Schwinger fields $\Psi_{\mu}^{i}$,
$i \in \lbrace 1,2,3 \rbrace$  \cite{Rarita:1941mf}. The LO Lagrangian reads 
\begin{align}
\mathcal{L}^{(1)}_{\pi \Delta} &= - \bar{\Psi}_{\mu}^{i} \xi_{ i j }^{3/2} \left \lbrace \left( i \slashed{D}^{ j k } - m_{\Delta 0}^{\phantom{()}}
\delta^{ j k } \right) g^{\mu \nu} - i \left( \gamma^{\mu} D^{\nu , \, j k } + \gamma^{\nu} D^{\mu , \, j k } \right) + i \gamma^{\mu} \slashed{D}^{ j k }
\gamma^{\nu}  \right. \notag\\  
&\phantom{= - \bar{\Psi}_{\mu}^{i} \xi_{ i j }^{3/2} } \quad \left. + m_{\Delta 0}^{\phantom{()}} \delta^{ j k } \gamma^{\mu} \gamma^{\nu}
+ \frac{g_1}{2} \slashed{u}^{ j k } \gamma_5 g^{\mu \nu} + \frac{g_2}{2} \left( \gamma^{\mu} u^{\nu , \, j k } + u^{\mu , \, j k } \gamma^{\nu} \right)
\gamma_5 \right. \notag \\ 
&\phantom{= - \bar{\Psi}_{\mu}^{i} \xi_{ i j }^{3/2} } \quad \left.+  \frac{g_3}{2} \gamma^{\mu} \slashed{u}^{ j k } \gamma_5 \gamma^{\nu}
\right \rbrace \xi_{ k l }^{3/2} \Psi_{\nu}^{l}  \; ,
\end{align}
where $m_{\Delta 0}$ is the chiral limit mass of the delta, $g_1$, $g_2$, and $g_3$ are coupling constants, that are, however,
not independent~\cite{Hacker:2005fh}. Further, $\xi_{ i j }^{3/2}$ is the isospin-$3/2$ projector  
\begin{align}
	\xi_{ i j }^{3/2} = \delta_{ij} - \frac{1}{3} \tau_i \tau_j \; , 
\end{align}
in terms of the Pauli-matrices $\tau_i$.
The propagator  $G^{\rho \mu} (k)$ of the spin-$3/2$ Rarita-Schwinger propagator in $D$ space-time dimensions is given by 
\begin{align}
\label{deltaprop}  
G^{\rho \mu} (k)  =  \frac{-i (\slashed{k} + m_{\Delta}^{} )}{k^2 - m_{\Delta}^2 + i \epsilon} \left( g^{ \rho \mu }
- \frac{1}{D-1} \gamma^{\rho} \gamma^{\mu} + \frac{k^{\rho}  \gamma^{\mu} -  \gamma^{\rho} k^{\mu}}{(D-1) m_{\Delta}^{} }
- \frac{D-2}{(D-1) m_{\Delta}^{2} } k^{\rho} k^{\mu} \right) \; ,
\end{align}
where we use the physical delta mass $m_{\Delta}$, which is legitimate in our calculation.
The  LO interactions between pions, nucleons, deltas and Roper resonances are completed by
\begin{align}
\mathcal{L}^{(1)}_{\pi N R} &= \bar{\Psi}_{R}^{\phantom{()}} \left( \frac{g_{\pi N R}}{2} \slashed{u} \gamma_5 \right) \Psi_{N}^{\phantom{()}}
+ \text{h.c.}  \; , \notag\\
\mathcal{L}^{(1)}_{\pi N \Delta} &= h \bar{\Psi}_{\mu}^{i} \xi_{ i j }^{3/2} \Theta^{\mu \alpha} \left( z_1 \right) \omega_{\alpha}^{j}
\Psi_{N}^{\phantom{()}} + \text{h.c.}  \; , \notag\\
\mathcal{L}^{(1)}_{\pi \Delta R} &= h_{R}^{\phantom{()}} \bar{\Psi}_{\mu}^{i} \xi_{ i j }^{3/2} \Theta^{\mu \alpha} \left( {z_2} \right)
\omega_{\alpha}^{j} \Psi_{R}^{\phantom{()}} + \text{h.c.}  \; . 
\end{align}
Here, $g_{\pi N R}$, $h$, and $h_R$ are coupling constants and 
\begin{align}
\omega_{\nu}^{j} = \frac{1}{2} \text{Tr} \left( \tau^j u_{\nu} \right) \; , \quad \Theta^{\mu \nu} \left( z \right) = g^{\mu \nu}
+ z \gamma^{\mu} \gamma^{\nu} \; ,
\end{align}
where $ z_1$ and ${z}_2$ are off-shell parameters. Throughout this text we follow Ref.~\cite{Gegelia:2016xcw} and set
$g_1 = - g_2 = - g_3$ and $ z_1 = {z}_2 =0 $, see also Refs.~\cite{Tang:1996sq,Krebs:2009bf}.
More terms have to be taken into account if one is interested in performing calculations of higher chiral order.

\section{Self-energy of the Roper resonance}
\label{sec:ropercont}
To calculate the mass of the Roper resonance in the infinite (and also finite) volume we have to determine the poles of the
dressed propagator
\begin{align}
	i S_{R} \left( p \right) = \frac{i}{\slashed{p} - m_{R 0} - \Sigma_{R} \left( \slashed{p} \right) } \; . 
\end{align}
Here, $\Sigma_{R}$ denotes the self-energy of the Roper, which can be calculated from all one-particle-irreducible
contributions to the two-point function of the Roper resonance field $\Psi_{R}$. The poles are  obtained by solving the equation
\begin{align}
\left. \left[ \slashed{p} - m_{R 0} - \Sigma_{R} \left( \slashed{p} \right) \right] \right|_{\slashed{p}={Z}} \overset{!}{=} 0\; , 
\end{align}
where in the infinite volume $Z$ is parametrized by 
\begin{align}
	Z  = m_R - i \frac{\Gamma_{R}}{2} \; ,
\end{align}
in terms of the physical Roper mass $m_R$ and its width $\Gamma_{R}$. This implies that the real part of the self-energy
corresponds to corrections of $m_R$, whereas the imaginary part corresponds to corrections of $\Gamma_{R}$.

\begin{figure}[t]
	\begin{center}  
		\includegraphics[width=0.70\linewidth]{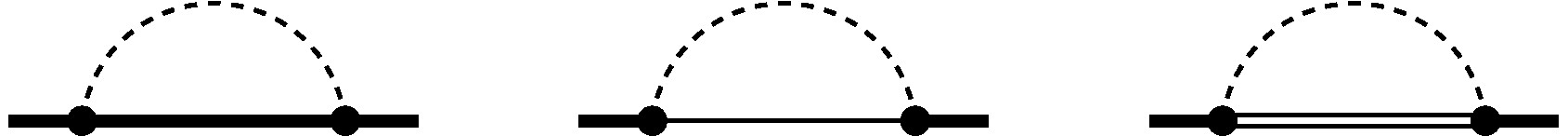}
	\end{center}
	\caption{
		One-loop diagrams contributing to the Roper mass at third chiral order. 
		Thick solid, dashed, solid, and solid double lines refer to the Roper resonance, pions, nucleons, and delta
                baryon states, respectively. The vertices denoted by a filled dot refer to insertions from the first order
                chiral Lagrangian.
	} 
	\label{fig:diags1}
\end{figure}

At third chiral order there are three one-loop diagrams contributing to the self-energy of the Roper resonance,
which are depicted in Fig.~\ref{fig:diags1}.
The diagrams differ by the internal baryon state, which can be a Roper, a nucleon or a delta baryon. Additionally, there is a
contact interaction coming from the second order Lagrangian of the Roper in Eq.~(\ref{PiRNLagr}). The self-energy up to order
$\mathcal{O} \left( p^3 \right)$ then reads
\begin{align}
\Sigma_{R}^{\phantom{()}} \left( \slashed{p} \right) = \underbrace{\Sigma_{R}^{(2)} \left( \slashed{p} \right)}_{\text{contact int.}}
+ \underbrace{ \Sigma_{ \pi R}^{(3)} \left( \slashed{p} \right) + \Sigma_{ \pi N}^{(3)} \left( \slashed{p} \right)  +
\Sigma_{ \pi \Delta}^{(3)} \left( \slashed{p} \right) }_{\text{loops}} + \mathcal{O} \left( p^4 \right) \; .
\end{align}
Using the effective Lagrangians from Sec.~\ref{SecEffLagr}, we can straightforwardly write down the expressions for
the self-energy. For the second order contact interaction we find
\begin{align}
	\Sigma_{R}^{(2)} = -4 c_{1}^R M_{\pi}^2 \; ,
\end{align}
and the three loop contributions are given by
\begin{align}
\Sigma_{ \pi R}^{(3)} \left( \slashed{p} \right) &= \frac{3 g_{R}^2}{4 F_{\pi}^2}  \int \frac{d^4 k }{(2 \pi)^4}
\frac{ i \slashed{k} \gamma_5 \left( \slashed{p} - \slashed{k} + m_R \right) \slashed{k} \gamma_5 }{ \lbrack ( p-k )_{\phantom{R}}^2
- m_R^2 + i \epsilon \rbrack \lbrack k_{\phantom{\pi}}^2 - M_{\pi}^2 + i \epsilon \rbrack } \; , \label{SOp3Roper} \\
\Sigma_{ \pi N}^{(3)} \left( \slashed{p} \right) &= \frac{3 g_{\pi N R}^2}{4 F_{\pi}^2}  \int \frac{d^4 k }{(2 \pi)^4} \frac{ i \slashed{k}
\gamma_5 \left( \slashed{p} - \slashed{k} + m_N \right) \slashed{k} \gamma_5 }{ \lbrack ( p-k )_{\phantom{N}}^2 - m_N^2 + i \epsilon \rbrack
\lbrack k_{\phantom{\pi}}^2 - M_{\pi}^2 + i \epsilon \rbrack } \; , \label{SOp3Nuc} \\
\Sigma_{ \pi \Delta}^{(3)} \left( \slashed{p} \right) &= \frac{2 h_{R}^2}{ F_{\pi}^2} \int \frac{d^4 k }{(2 \pi)^4} \frac{\left( p-k \right)_{\mu}
G^{\mu \nu} \left( k \right) \left( p-k \right)_{\nu}}{( p-k )_{\phantom{\pi}}^2 - M_{\pi}^2 + i \epsilon} \; , \label{SOp3Delta}
\end{align}
where  $G^{\rho \mu} (k)$ is given  in Eq.~(\ref{deltaprop}).
The three one-loop contributions to the Roper mass can be expanded in terms of the scalar Passarino-Veltman integrals (PV integrals).
This expansion is done using the Mathematica package FeynCalc \cite{Mertig:1990an,Shtabovenko:2016sxi}. 
The definitions and solutions of the PV integrals can be found in App.~\ref{AppPVIntegrals}.
This results in 
\begin{align}
\Sigma_{ \pi R}^{(3)} \left( \slashed{p} = m_{R}^{} \right) &= \frac{3 g_{R}^2}{4 F_{\pi}^2} \left. \int \frac{d^4 k }{(2 \pi)^4} \frac{ i \slashed{k} \gamma_5 \left( \slashed{p} - \slashed{k} + m_R \right) \slashed{k} \gamma_5 }{ \lbrack ( p-k )_{\phantom{R}}^2 - m_R^2 + i \epsilon \rbrack \lbrack k_{\phantom{\pi}}^2 - M_{\pi}^2 + i \epsilon \rbrack } \right|_{\slashed{p} = m_{R}^{} } \notag \\
&= \frac{3 g_{R}^2 m_{R}^{} }{32 F_{\pi}^2} \left \lbrace M_{\pi}^2 B_0 \left( m_R^2, m_R^2, M_{\pi}^2 \right) + A_0 \left( m_R^2 \right) \right \rbrace \; , \\
& \notag \\
\Sigma_{ \pi \Delta}^{(3)} \left( \slashed{p} = m_{R}^{} \right) &= \frac{2 h_{R}^2}{F_{\pi}^2}  \left. \int \frac{d^4 k }{(2 \pi)^4} \frac{\left( p-k \right)_{\mu} G^{\mu \nu} \left( k \right) \left( p-k \right)_{\nu}}{( p-k )_{\phantom{\pi}}^2 - M_{\pi}^2 + i \epsilon} \right|_{\slashed{p} = m_{R}^{} } \notag \\ 
&= \frac{h_{R}^2}{96 \pi^2 F_{\pi}^2 m_{\Delta}^2 m_{R}^{} } \left \lbrace \phantom{\left( M_{\pi}^2 \right)^2}  \right. \notag \\ 
& \left. \phantom{\left( M_{\pi}^2 \right)^2}  - \left[ \left( m_{\Delta}^{} + m_{R}^{} \right)^2 - M_{\pi}^2 \right] \lambda \left( m_{R}^{2}, m_{\Delta}^{2}, M_{\pi}^{2} \right) B_0 \left( m_{R}^{2}, m_{\Delta}^2, M_{\pi}^2 \right) \right. \notag \\ 
& \left. \phantom{\left( M_{\pi}^2 \right)^2} + \left[ m_{\Delta}^4 + 2 m_{\Delta}^3 m_{R}^{} + \left( M_{\pi}^2 - m_{R}^{2} \right)^2 - m_{\Delta}^2 \left( m_{R}^{2} + 2 M_{\pi}^2 \right) \right. \right. \notag \\
& \left. \left. \phantom{\left( M_{\pi}^2 \right)^2} + 2 m_{\Delta}^{} m_{R}^{} \left( m_{R}^{2} - M_{\pi}^2 \right) \right] A_0 \left( m_{\Delta}^2 \right) + \left[ m_{\Delta}^4 + 2 m_{\Delta}^3 m_{R}^{} - 2 m_{\Delta}^{} m_{R}^{3}  \phantom{\left( M_{\pi}^2 \right)^2} \right. \right. \notag \\ 
&\left. \left. \phantom{\left( M_{\pi}^2 \right)^2} - m_{R}^{4} - 2 m_{\Delta}^2 M_{\pi}^2 + 6 m_{\Delta}^{} m_{R}^{} M_{\pi}^2 + 5 m_{R}^{2} M_{\pi}^2 + M_{\pi}^4 \right] A_0 \left( M_{\pi}^2 \right) \right \rbrace \notag \\
& \phantom{=} \; + \frac{h_{R}^2}{576 \pi^2 F_{\pi}^2 m_{\Delta}^2 } \left \lbrace 3 m_{\Delta}^{4} m_{R}^{} - 12 m_{\Delta}^{3} m_{R}^{2} - 4 m_{\Delta}^{2} m_R^3 + 2 m_R^5 \phantom{\left( M_{\pi}^2 \right)^2} \right. \notag \\ 
& \left. \phantom{\left( M_{\pi}^2 \right)^2} -8 m_{R}^{3} M_{\pi}^{2} + 13 m_{R}^{} M_{\pi}^{4} + 4 m_{\Delta}^{}  \left( m_{R}^{4} - 3 m_{R}^{2} M_{\pi}^{2} + 4 M_{\pi}^{4} \right) \right \rbrace \; , 
\end{align}
\begin{align}
	\Sigma_{ \pi N}^{(3)} \left( \slashed{p} = m_{R}^{} \right) &= \frac{3 g_{\pi N R}^2}{4 F_{\pi}^2} \left. \int \frac{d^4 k }{(2 \pi)^4} \frac{ i \slashed{k} \gamma_5 \left( \slashed{p} - \slashed{k} + m_N \right) \slashed{k} \gamma_5 }{ \lbrack ( p-k )_{\phantom{N}}^2 - m_N^2 + i \epsilon \rbrack \lbrack k_{\phantom{\pi}}^2 - M_{\pi}^2 + i \epsilon \rbrack } \right|_{\slashed{p} = m_{R}^{} } \notag \\ 
	&= \frac{- 3 g_{\pi N R}^2}{128 \pi^2 F_{\pi}^2 m_{R}^{} } \left \lbrace \left( m_{R}^{} + m_N^{} \right)^2 \left[ \left( \left( m_{R}^{} - m_N^{} \right)^2 - M_{\pi}^2 \right) B_0 \left( m_{R}^{2}, m_N^2, M_{\pi}^2 \right) \right. \right. \notag \\ 
	& \left. \left. \phantom{\left( M_{\pi}^2 \right)^2} - A_0 \left( m_N^2 \right) \right] - \left( m_R^{2} - m_{N}^{2} \right) A_0 \left( M_{\pi}^2 \right) \right \rbrace \; .
\end{align}
We used K\"all\'en's triangle function $\lambda (x,y,z) = (x-y-z)^2 - 4 yz$ to simplify the lengthy expression
of $\Sigma_{ \pi \Delta}^{(3)}$. 

Evaluating these scalar integrals in the infinite volume leads to the well-known infinities that one has to tame within
the framework of renormalization. Procedures like the $\widetilde{\text{MS}}$ scheme use redefinitions of the bare
parameters in the Lagrangian to subtract the infinities. Additionally in baryonic ChPT one will encounter terms in the
expansion of the loop diagrams that break the power counting. These so-called power counting violating terms can be
handeld with different techniques, like the heavy baryon approach,  IR or the
EOMS scheme. Within this EOMS scheme one performs additionally finite subtractions to cancel the power counting violating
terms. In the end one obtains a finite result that is consistent with the power counting. Further details relevant for our
calculations can be found, e.g., in Refs.~\cite{Fuchs:2003qc,Severt:2019sbz}.

\section{Finite volume formalism}
\label{sec:finvol}
Next, we consider the Roper resonance in a finite volume. We place our system in a cubic box of length $L$ and calculate the
difference between the finite and infinite volume case~\cite{Luscher:1990ux}. In the finite volume the
(Euclidean) loop integral is replaced by an
infinite sum of the spatial momenta while the integration over the time component remains unchanged (in actual lattice QCD
calculations, the time direction is also discrete, but we keep it continuous for simplicity)
\begin{align}
  \int \frac{d^4 k_E}{(2 \pi)^4} \left(\ldots \right) \mapsto \int_{-\infty}^{\infty} \frac{d k_4}{2 \pi} \frac{1}{L^3} \sum_{\vec{k}}
  \left(\ldots \right) \; .
\end{align}
In a finite volume the spatial momenta are discretized and can only take  values that are integer multiples of $2 \pi / L$, i.e. 
(for a general discussion of theories with spontaneous symmetry breaking in a finite volume, see e.g. Ref.~\cite{Gasser:1987zq})
\begin{align}
	\vec{k} =  \frac{2 \pi}{L} \vec{n} \; , \quad \vec{n} \in \mathbb{Z}^3 \; . \label{discretemomenta}
\end{align}
This change obviously influences the self-energy of the Roper resonance. 
The poles of the propagator are now given by 
\begin{align}
\slashed{p} - m_{R0}^{}  - \Sigma_{R}^L \left( \slashed{p} \right) = 0 \; , \label{FVpoles}
\end{align}
where $\Sigma_{R}^L \left( \slashed{p} \right)$ denotes the self-energy of the Roper in the finite box. The difference beetween the
self-energy in the box and in the infinite volume is defined as the finite volume correction (FV correction) of the system
\cite{AliKhan:2003ack,Beane:2004rf}
\begin{align}
  \tilde{\Sigma}_{R}^L \left( \slashed{p} \right) := \Sigma_{R}^L \left( \slashed{p} \right)
  - \text{Re} \left \lbrace \Sigma_{R}^{} \left( \slashed{p} \right) \right \rbrace \; . \label{FVcorrec}
\end{align}
Note that in the finite volume the self-energy can only yield real values due to the summation over real momenta $\vec{k}$.
Therefore we have to restrict the infinite volume self-energy to its real part to ensure a non-imaginary FV correction. 

Using Eq.~(\ref{FVcorrec}) we can reformulate Eq.~(\ref{FVpoles}). We choose the center-of-mass frame $p_{\mu} = ( E, \vec{0} )$ and
use the on-shell condition $\slashed{p} = E$ to obtain  
\begin{align}
  0 & \overset{!}{=} E - m_{R 0}^{}  - \Bigl\lbrack \tilde{\Sigma}_{R}^L \left( E \right) + \text{Re} \left \lbrace \Sigma_{R}^{}
  \left( E \right) \right \rbrace \Bigr\rbrack \notag \\
  & = E - \underbrace{\Bigl\lbrack m_{R 0}^{} + \text{Re} \left \lbrace \Sigma_{R}^{} \left( E \right) \right \rbrace \Bigr\rbrack}_{ m_R^{} }
  - \tilde{\Sigma}_{R}^L \left( E \right) \quad \Leftrightarrow \quad  m_{R}^{} - E = - \tilde{\Sigma}_{R}^L \left( E \right) \; , \notag 
\end{align}
where we used the definition of the physical Roper mass $m_R$, i.e. the real part of the pole, in the last step.
At third chiral order, the contributions to the self-energy are the ones from Fig.~\ref{fig:diags1}. We get 
\begin{align}
  m_{R}^{} - E = - \left \lbrace \tilde{\Sigma}_{\pi R}^{L, \, (3)} \left( E \right) + \tilde{\Sigma}_{\pi N}^{L, \, (3)} \left( E \right)
  + \tilde{\Sigma}_{\pi \Delta}^{L, \, (3)} \left( E \right) \right \rbrace \; , \label{ENLevelsEq1}
\end{align}
and our goal will be the numerical evaluation of this equation. The three one-loop contributions to the Roper mass have been expanded
in terms of the PV integrals in the last section. Now we have to replace the infinite volume quantities with their finite volume
expressions. We obtain 
\begin{align}
\tilde{\Sigma}_{\pi R}^{L, \, (3)} \left( E \right) &= - \frac{3 g_R^2}{128 \pi^2 F_{\pi}^2 E} \left( E+ m_R^{} \right) \left
\lbrace \left( E+ m_R^{} \right) \left[ \left( \left( E - m_R^{} \right)^2 - M_{\pi}^2 \right) \tilde{B}_0^L
\left( E^2, m_R^2, M_{\pi}^2 \right) \right. \right. \notag \\ 
& \left. \left. \phantom{\left( M_{\pi}^2 \right)^2} - \tilde{A}_0^L \left( m_R^2 \right) \right] + \left( m_R^{} -E \right)
\tilde{A}_0^L \left( M_{\pi}^2 \right) \right \rbrace \; , 
\end{align}
\begin{align}
\tilde{\Sigma}_{\pi N}^{L, \, (3)} \left( E \right) &= - \frac{3 g_{\pi N R}^2}{128 \pi^2 F_{\pi}^2 E} \left( E+ m_N^{} \right)
\left \lbrace \left( E+ m_N^{} \right) \left[ \left( \left( E - m_N^{} \right)^2 - M_{\pi}^2 \right) \tilde{B}_0^L
\left( E^2, m_N^2, M_{\pi}^2 \right) \right. \right. \notag \\ 
& \left. \left. \phantom{\left( M_{\pi}^2 \right)^2} - \tilde{A}_0^L \left( m_N^2 \right) \right] + \left( m_N^{} -E \right)
\tilde{A}_0^L \left( M_{\pi}^2 \right) \right \rbrace \; ,
\end{align}
\begin{align}
\tilde{\Sigma}_{\pi \Delta}^{L, \, (3)} \left( E \right) &= \frac{h_{R}^2}{96 \pi^2 F_{\pi}^2 m_{\Delta}^2 E} \left \lbrace -
\left[ \left( m_{\Delta}^{} + E \right)^2 - M_{\pi}^2 \right] \lambda \left( E^2, m_{\Delta}^{2}, M_{\pi}^{2} \right) \tilde{B}_0^L
\left( E^2, m_{\Delta}^2, M_{\pi}^2 \right) \right. \notag \\ 
& \left. \phantom{\left( M_{\pi}^2 \right)^2} + \left[ m_{\Delta}^4 + 2 m_{\Delta}^3 E + \left( M_{\pi}^2 - E^2 \right)^2 -
m_{\Delta}^2 \left( E^2 + 2 M_{\pi}^2 \right) \right. \right. \notag \\
& \left. \left. \phantom{\left( M_{\pi}^2 \right)^2} + 2 m_{\Delta}^{} E \left( E^2 - M_{\pi}^2 \right) \right] \tilde{A}_0^L
\left( m_{\Delta}^2 \right) + \left[ m_{\Delta}^4 + 2 m_{\Delta}^3 E - 2 m_{\Delta}^{} E^3  \phantom{\left( M_{\pi}^2 \right)^2}
\right. \right. \notag \\ 
&\left. \left. \phantom{\left( M_{\pi}^2 \right)^2} - E^4 - 2 m_{\Delta}^2 M_{\pi}^2 + 6 m_{\Delta}^{} E M_{\pi}^2 + 5 E^2 M_{\pi}^2
+ M_{\pi}^4 \right] \tilde{A}_0^L \left( M_{\pi}^2 \right) \right \rbrace \; ,
\end{align}
where $\tilde{A}_0^L$ and $\tilde{B}_0^L$ are the finite volume corrections of the PV integrals which will be determined next.

\subsection{Calculation of loop integrals in the finite volume}
Let us consider as an example 
\begin{align}
	A_0 (m^2) = -16\pi^2  \int \frac{d^4 k}{(2 \pi)^{4}} \frac{i}{k^2 - m^2} 
\end{align}
in four-dimensional Minkowski space. We will follow the procedure described in Ref.~\cite{Bernard:2007cm} here.
First of all we perform the Wick rotation $k_0 \to i k_4$ to Euclidean space, so that the integral can be rewritten as 
\begin{align}
  A_0 \left( m^2 \right) &
  = -16\pi^2   \int \frac{d^4 k_E}{(2 \pi)^{4}} \frac{1}{k_{E}^2 + m_{\phantom{4}}^2 } \; . \label{PVinfEuclid}
\end{align}
Now we can define the finite volume PV integral by replacing the Euclidean spatial integral with a discrete sum 
\begin{align}
  A_0^{L} \left( m^2 \right) = -16\pi^2  \int_{-\infty}^{\infty} \frac{d k_4}{2 \pi} \frac{1}{L^3} \sum_{\vec{k}}
  \frac{1}{k_{4}^2 + | \vec{k} |_{\phantom{4}}^2 + m_{\phantom{4}}^2 } \; , 
\end{align}
where the momenta $\vec{k}$ are restricted according to Eq.~(\ref{discretemomenta}). The evaluation of this sum is the next task. First,
we note that the function inside the sum is regular, i.e. it does not possess a pole on the real axis for all values of $k_4$ and $\vec{k}$.
Therefore we can use the so called Poisson trick to simplify the calculation. We insert the Dirac delta into the equation 
\begin{align}
  A_0^{L} \left( m^2 \right) = -16\pi^2   \int_{-\infty}^{\infty} \frac{d k_4}{2 \pi} \frac{1}{L^3} \int d^3 k
  \frac{1}{k_{4}^2 + | \vec{k} |_{\phantom{4}}^2 + m_{\phantom{4}}^2 } \sum_{\vec{n}} \delta^{(3)} \left( \vec{k} - \frac{2 \pi}{L} \vec{n} \right) \; , 
\end{align}
and then we use the Poisson formula in three dimensions
\begin{align}
  \sum_{\vec{n}} \delta^{(3)} \left( \vec{k} - \frac{2 \pi}{L} \vec{n} \right) &
	= \left( \frac{L}{2 \pi} \right)^3 \sum_{\vec{n}} \exp \left(  i L \vec{n} \cdot \vec{k} \right) \; .
\end{align}
Plugging this result in our finite volume PV integral we obtain 
\begin{align}
	A_0^{L} \left( m^2 \right) = -16\pi^2 \int \frac{d^4 k_E}{(2 \pi)^4} \frac{1}{k_{4}^2 + | \vec{k} |_{\phantom{4}}^2 + m_{\phantom{4}}^2 } \sum_{\vec{n}} e^{i L \vec{k} \cdot \vec{n} } \; , 
\end{align}
where we have regained a four-dimensional integral over momenta times a sum of exponential functions. We observe that the term in
the sum with $\vec{n} = \vec{0}$ reproduces the infinite volume PV integral from Eq.~(\ref{PVinfEuclid}) 
\begin{align}
A_0^{L} \left( m^2 \right) &= -16\pi^2 \left \lbrace \int \frac{d^4 k_E}{(2 \pi)^4} \frac{1}{k_{4}^2
  + | \vec{k} |_{\phantom{4}}^2 + m_{\phantom{4}}^2 } + \sum_{\vec{n} \neq 0} \int \frac{d^4 k_E}{(2 \pi)^4}
\frac{e^{i L \vec{k} \cdot \vec{n} }}{k_{4}^2 + | \vec{k} |_{\phantom{4}}^2 + m_{\phantom{4}}^2 } \right \rbrace \notag \\
&= A_0^{} \left( m^2 \right) - 16 \pi^2 \sum_{\vec{n} \neq 0} \int \frac{d^4 k_E}{(2 \pi)^4}
\frac{e^{i L \vec{k} \cdot \vec{n} }}{k_{4}^2 + | \vec{k} |_{\phantom{4}}^2 + m_{\phantom{4}}^2 } \; .
\end{align}
Thus, the finite volume correction is given by 
\begin{align}
  \tilde{A}_0^{L} \left( m^2 \right) := A_0^{L} \left( m^2 \right) - A_0^{} \left( m^2 \right) = - 16 \pi^2 \sum_{\vec{n} \neq 0} \int
  \frac{d^4 k_E}{(2 \pi)^4} \frac{e^{i L \vec{k} \cdot \vec{n} }}{k_{4}^2 + | \vec{k} |_{\phantom{4}}^2 + m_{\phantom{4}}^2 } \; . 
\end{align}
The remaining integral is finite and can be solved with standard methods. After integrating the spatial part we are left with 
\begin{align}
  \tilde{A}_0^{L} \left( m^2 \right) = - 4 \sum_{j \neq 0} \frac{1}{L j} \int_{0}^{\infty} d k_4 e^{- L j \sqrt{k_{4}^2 + m_{\phantom{4}}^2}}
  = -4 m^2 \sum_{j \neq 0} \frac{K_1 \left( m L j  \right)}{m L j} \; , 
\end{align}
where $j = | \vec{n} | =\sqrt{ \strut n_1^2 + n_2^2 + n_3^2} $ and $K_{\nu} (z)$ is the modified Bessel function of the second kind.
For large values of the box length $L$ and the summation index $j$ the finite volume correction decreases exponentially
so that it becomes negligible for large volumes (usually this is expected for $M_\pi L > 4$).

A similar calculation can be done for $\tilde{B}_0^{L}$. After performing the Wick rotation in the infinite volume, the
integral has the form 
\begin{align}
B_0^{} \left( E^2, m_X^2, M_{\pi}^2 \right) = 16 \pi^2 \int \frac{d^4 k_E}{(2 \pi)^4} \frac{1}{\lbrack k_{E}^2 + M_{\pi}^2 \rbrack
\lbrack ( \hat{P} - k_E )_{\phantom{X}}^2 + m_X^2 \rbrack} \; , \quad  \hat{P}_{\mu} = ( i E , \vec{0} ) \; ,
\end{align}
and the finite volume expression is given by 
\begin{align}
B_0^{L} \left( E^2, m_X^2, M_{\pi}^2 \right) = 16 \pi^2 \int_{-\infty}^{\infty} \frac{d k_4}{2 \pi} \frac{1}{L^3} \sum_{\vec{k}}
\frac{1}{\lbrack k_{4}^2 + | \vec{k} |_{\vphantom{4}}^2 + M_{\pi}^2 \rbrack  \lbrack ( i E - k_4^{} )_{\vphantom{X}}^2 + | \vec{k} |_{\vphantom{X}}^2
 + m_X^2 \rbrack} \; . 
\end{align}
The next step is to use Feynman parameterization, see App.~\ref{AppUsefformula} for further details, to combine the two
denominators and then perform a shift in the non-discret momentum component $k_4$. The resulting expression reads  
\begin{align}
B_0^{L} \left( E^2, m_X^2, M_{\pi}^2 \right) = 16 \pi^2 \int_{0}^{1} dy \int_{-\infty}^{\infty} \frac{d k_4}{2 \pi}
\frac{1}{L^3} \sum_{\vec{k}}
\frac{1}{\left[ k_{4}^2 + | \vec{k} |_{\vphantom{4}}^2 + g_{X} \left( y, E^2 \right) \right]^2 } \; ,
\end{align}
with
\begin{align}
g_{X} \left( y, E^2 \right) = y \left( y-1  \right) E^2 + y m_X^2 + \left( 1-y \right) M_{\pi}^2 \; .
\end{align}
Depending on the values for $E$, $m_X$ and $M_{\pi}$, the function $g_{X} \left( y, E^2 \right)$ can be positive, negative or
zero for some values of $y$. If $g_{X} \left( y, E^2 \right) > 0$ for all $y \in \left[ 0,1 \right] $, the function inside
the sum is regular and we can again use the Poisson formula analogously to $A_0^L$. The finite volume correction then is 
\begin{align}
  \tilde{B}_0^{L} \left( E^2, m_X^2, M_{\pi}^2 \right) = 2 \int_{0}^{1} dy \sum_{j \neq 0} K_{0} \left( L j \sqrt{\strut g_{X}^{}
  \left( y, E^2 \right)} \right) \; .
\end{align}
Also here the correction drops exponentially for large $L$ and $j$. Note that the parameter integral over $y$ has to be
evaluated numerically.

However, if the function $g_X$ is negative or equal to zero for some values of $y$, the Poisson formula is no longer applicable.
In our calculation for example, if $m_X = m_R$ the difference between the pole position $E$ and $m_R$ is small and $g_X$
stays positive. If $m_X = m_N, \, m_{\Delta}$ the function can become negative and we have to find another way to evaluate
the finite volume contribution.
To do so we follow again Ref.~\cite{Bernard:2007cm} and introduce a scale $\mu$, which will be used to subtract ultraviolet
divergences from the infinite sum. Also the scale can be choosen in such a way that the function $g_X \left( y, \mu^2 \right)$
stays positive. We expand the finite volume correction as 
\begin{align}
\tilde{B}_0^{L} \left( E^2, m_X^2, M_{\pi}^2 \right) &= B_0^{L} \left( E^2, m_X^2, M_{\pi}^2 \right) -  \text{Re} \left \lbrace B_0^{} \left( E^2, m_X^2, M_{\pi}^2 \right) \right
\rbrace \notag \\ 
&= B_0^{L} \left( E^2, m_X^2, M_{\pi}^2 \right) -  \text{Re} \left \lbrace B_0^{} \left( E^2, m_X^2, M_{\pi}^2 \right) \right
\rbrace \notag \\ 
& \phantom{=} \; + \tilde{B}_0^{L} \left( \mu^2, m_X^2, M_{\pi}^2 \right) - B_0^{L} \left( \mu^2, m_X^2, M_{\pi}^2 \right) + B_0^{} \left( \mu^2, m_X^2, M_{\pi}^2 \right) \notag \\
& \phantom{=} \; + \left( E^2 - \mu^2 \right) \frac{d}{d E^2} \Big\{ \tilde{B}_0^{L} \left( E^2, m_X^2, M_{\pi}^2 \right) \phantom{\tilde{B}_0^{L}}  \notag \\
& \phantom{=} \;  - B_0^{L} \left( E^2, m_X^2, M_{\pi}^2 \right) + B_0^{} \left( E^2, m_X^2, M_{\pi}^2 \right) \Big\} \bigg|_{E^2 = \mu^2} \notag \\
& \equiv 16 \pi^2 \left \lbrace H_1^{X} \left( E^2 \right) + H_2^{X} \left( E^2 \right) + H_3^{X} \left( E^2 \right) \right \rbrace \; , 
\end{align}
and separate it into three terms, which are given by 
\begin{align}
	16\pi^2 \, H_1^{X} \left( E^2 \right) &= \bigg\{ B_0^{L} \left( E^2, m_X^2, M_{\pi}^2 \right) - B_0^{L} \left( \mu^2, m_X^2, M_{\pi}^2 \right)  \notag \\
	& \phantom{= = =} \; - \left( E^2 - \mu^2 \right) \frac{d}{d E^2} B_0^{L} \left( E^2, m_X^2, M_{\pi}^2 \right) \bigg|_{E^2 = \mu^2} \bigg\} \; ,  \\
	16\pi^2 \, H_2^{X} \left( E^2 \right) &= \bigg\{ \tilde{B}_0^{L} \left( \mu^2, m_X^2, M_{\pi}^2 \right) + \left( E^2 - \mu^2 \right) \frac{d}{d E^2} \tilde{B}_0^{L} \left( E^2, m_X^2, M_{\pi}^2 \right) \bigg\} \; , \\
	16\pi^2 \, H_3^{X} \left( E^2 \right) &= - \bigg\{ \text{Re} \left \lbrace B_0^{} \left( E^2, m_X^2, M_{\pi}^2 \right) \right \rbrace - B_0^{} \left( \mu^2, m_X^2, M_{\pi}^2 \right)  \notag \\
	& \phantom{= = = - } \; - \left( E^2 - \mu^2 \right) \frac{d}{d E^2} B_0^{} \left( E^2, m_X^2, M_{\pi}^2 \right) \bigg|_{E^2 = \mu^2} \bigg\} \; .
\end{align}
The first subtraction term with the newly introduced scale $\mu$ ensures that the correction converges, while the derivative terms
lead to a faster convergence. The first term, $H_1^X$, contains only terms with momentum sums. After the integration over $k_4$,
one obtains 
\begin{align}
	H_1^{X} \left( E^2 \right) = \left( E^2 - \mu^2 \right)^2 \frac{1}{L^3} \sum_{\vec{n}} \frac{E_X + E_{\pi} }{2 E_X E_{N} } \frac{1}{(E_X + E_{\pi})^2 - E^2} \frac{1}{((E_X + E_{\pi})^2 - \mu^2)^2} \; ,
\end{align}
with $E_X = \sqrt{m_X^2 + \left( \frac{2 \pi}{L} \right)^2 | \vec{n} |^2 }$ and $E_{\pi} = \sqrt{M_{\pi}^2 + \left( \frac{2 \pi}{L}
  \right)^2 |\vec{n} |^2 }$. In the second term one finds finite volume corrections that can be calculated with the Poisson
summation formula, since the functions are regular. We get  
\begin{align}
	H_2^{X} \left( E^2 \right) &= \frac{1}{8 \pi^2} \int_{0}^{1} dy \sum_{j \neq 0} \left \lbrace K_0 \left( L j \sqrt{g_X \left( y, \mu^2 \right)} \right)\phantom{ \frac{y (y-1) L j }{2 \sqrt{g_X \left( y, \mu^2 \right)}} }  \right. \notag \\
	& \phantom{= \frac{1}{8 \pi^2} \int_{0}^{1} dy } \left. - \left( E^2 - \mu^2 \right) \frac{y (y-1) L j }{2 \sqrt{g_X \left( y, \mu^2 \right)}} K_1 \left( L j \sqrt{g_X \left( y, \mu^2 \right)} \right) \right \rbrace \; . 
\end{align} 
The last term only contains infinite volume quantities which can be calculated with standard methods
\begin{align}
	H_3^{X} \left( E^2 \right) & = - \frac{B_{E}}{32 \pi^2 E^2} \left \lbrace \ln \left( \frac{E^2 + m_X^2 - M_{\pi}^2 + B_E }{E^2 + m_X^2 - M_{\pi}^2 - B_E } \right) + \ln \left( \frac{E^2 - m_X^2 + M_{\pi}^2 + B_E }{E^2 - m_X^2 + M_{\pi}^2 - B_E } \right) \right \rbrace \notag \\
	& \phantom{=} \; + \frac{B_{\mu}}{16 \pi^2 \mu^2} \left \lbrace \arctan \left( \frac{\mu^2 + m_X^2 - M_{\pi}^2 }{B_{\mu}} \right) + \arctan \left( \frac{\mu^2 - m_X^2 + M_{\pi}^2 }{B_{\mu}} \right) \right \rbrace \notag \\
	& \phantom{=} \; - \frac{E^2 - \mu^2}{16 \pi^2 \mu^2} \left \lbrace 1 + \frac{\left( E^2 - \mu^2 \right)\left( m_X^2 - M_{\pi}^2 \right)}{2 E^2 \mu^2} \ln \left( \frac{M_{\pi}^2}{m_X^2} \right)   \right. \notag \\
	& \phantom{=} \; \left. \phantom{ \frac{\left(M_{\pi}^2 - \mu^2\right)}{16 \pi^2 \mu^2} } + \frac{\left( m_X^2 - M_{\pi}^2 \right)^2 - \mu^2 \left( m_X^2 + M_{\pi}^2 \right)}{\mu^2 B_{\mu}}  \right. \notag \\
	& \phantom{=} \; \left. \phantom{ \frac{\left(M_{\pi}^2 - \mu^2\right)}{16 \pi^2 \mu^2} } \times \left[ \arctan \left( \frac{\mu^2 + m_X^2 - M_{\pi}^2 }{B_{\mu}} \right) + \arctan \left( \frac{\mu^2 - m_X^2 + M_{\pi}^2 }{B_{\mu}} \right) \right] \right \rbrace \; , 
\end{align}
where we used again the triangle function to define
\begin{align}
B_E &= \lambda^{1/2} \left( E^2, m_X^2, M_{\pi}^2 \right) := \sqrt{ \left( E^2 - m_X^2 - M_{\pi}^2 \right)^2 - 4 m_X^2 M_{\pi}^2 } \; , \\ 
B_{\mu} &= i \lambda^{1/2} \left( \mu^2, m_X^2, M_{\pi}^2 \right) := \sqrt{ - \left( \mu^2 - m_X^2 - M_{\pi}^2 \right)^2 + 4 m_X^2 M_{\pi}^2 } \; . 
\end{align}
We now have evaluated all PV integrals in the finite volume that we need. We note that the issue of using the PV reduction
in the finite volume was already discussed in Ref.~\cite{Bernard:2007cm} and we refer to that paper for details.
Thus we return to the main task, the
numerical calculation of Eq.~(\ref{ENLevelsEq1}).

\section{Results}
\label{sec:res}
\subsection{Calculation of the energy levels}
We now want to determine the energy spectrum of the Roper resonance system. To obtain this we take a look at
Eq.~(\ref{ENLevelsEq1}) and try to find numerical solutions for the energy $E$ for different box sizes $L$. Due to the
presence of the Roper we expect to see the so-called avoided level crossing of the energy levels.

Before we start to solve Eq.~(\ref{ENLevelsEq1}) by numerical methods, let us again consider the results of the finite
volume PV integrals. We have seen that all regular functions in the self-energy of the Roper decrease exponentially for
large $L$. This includes all tadpoles, i.e. all $\tilde{A}_0^L$ functions, as well as $\tilde{B}_0^L \left( E^2, m_R^2, M_{\pi}^2
\right)$ from the $\pi R$ loop in Fig.~\ref{fig:diags1}. Choosing $L$ to be large we can neglect the contributions from these
functions, leaving just the FV correction from the $\pi N$ and the $\pi \Delta$ loop (see also
Refs.~\cite{Bernard:2007cm,Bernard:2009mw}). This facilitates the numerical computation
of the energy levels significantly. The simplified equation reads 
\begin{align}
m_{R}^{} - E &= \frac{3 g_{\pi N R}^2}{128 \pi^2 F_{\pi}^2 E} \left( E+ m_N^{} \right)^2 \left[ \left( E - m_N^{} \right)^2 -
M_{\pi}^2 \right] \tilde{B}_0^L \left( E^2, m_N^2, M_{\pi}^2 \right) \notag \\ 
&\phantom{=} \; + \frac{h_{R}^2}{96 \pi^2 F_{\pi}^2 m_{\Delta}^2 E} \left[ \left( m_{\Delta}^{} + E \right)^2 - M_{\pi}^2 \right]
\lambda \left( E^2, m_{\Delta}^{2}, M_{\pi}^{2} \right) \tilde{B}_0^L \left( E^2, m_{\Delta}^2, M_{\pi}^2 \right) \; ,
\label{ENLevelsEq2} 
\end{align}
where only the non-regular functions and two LECs ($g_{\pi N R}$ and $h_R$) are left. 
Leaving out these contributions also simplifies the treatment of power counting breaking terms that would normally
appear in such a calculation. The remaining expression, however, does not contain any power counting violating terms,
so that an additional renormalization scheme, like EOMS, is redundant. Additional remarks on this issue are given in
Ref.~\cite{Bernard:2007cm}.  The further numerical studies of the energy levels are performed by using this equation.
Values of the used parameters are given in the next subsection.

\subsection{Numerical results}
For the hadron masses and constants we use the numerical values from Ref.~\cite{Gegelia:2016xcw}. The baryon masses are
$m_N = \SI{939}{\MeV}$, $m_R = \SI{1365}{\MeV}$~\footnote{This is the more reliable pole mass \cite{Tanabashi:2018oca}.},
and $m_{\Delta} = \SI{1210}{\MeV}$. For the pion mass we use
$M_{\pi} = \SI{139}{\MeV}$ and for the pion decay constant $F_{\pi} = \SI{92.2}{\MeV}$. The two coupling constants
are also taken from Ref.~\cite{Gegelia:2016xcw}
and are $g_{\pi N R} = \pm 0.47$, and $h_R = h = 1.42$, with the assumption that the coupling $h_R$ is equal to the
pion-nucleon-delta-coupling $h$ (the so-called maximal mixing \cite{Beane:2002ud}). Note that the sign of
$g_{\pi N R}$ does not matter
as this coupling appears squared in our analysis. We also have to choose the scale $\mu$ for
the calculation and set $\mu=m_N$ for the nucleon, and $\mu=m_{\Delta}$ for the delta case.

\begin{figure}[t]
	\begin{center}  
		\includegraphics[width=0.6\linewidth]{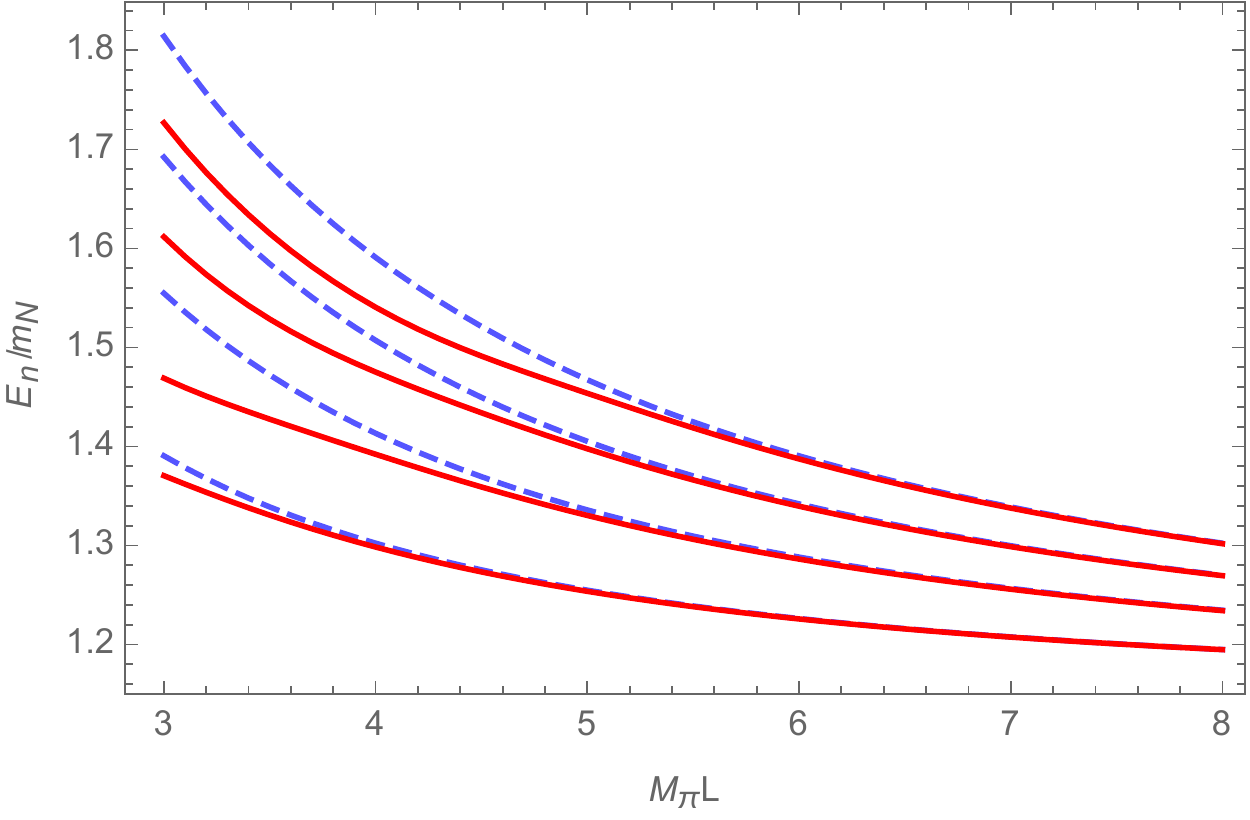}
	\end{center}
	\caption{
	  Energy levels for different box sizes $L$ considering only pion and nucleon as intermediate states. Red solid
          lines display the numerical results and blue dashed lines the free energy levels of the pion and nucleon
          for $|\vec{n}|^2 = 1,2,3,4$ (lowest to highest curve). 
	} 
	\label{fig:FitonlyNuc}
\end{figure}

Now we have everything we need to find the numerical values of $E$. We evaluate the sums in the finite volume corrections
from $|\vec{n}|^2 = 1$ up-to-and-including $|\vec{n}|^2 = 4$. Then we solve Eq.~(\ref{ENLevelsEq2}) for the respective energy
levels for different box sizes. To make things easier we first look at the Roper resonance without the delta, i.e. we
set $h_R=0$ and take only the interaction between Roper, nucleon and pion into account. 
The results are  displayed in Fig.~\ref{fig:FitonlyNuc}. 
The energy is depicted in units of the nucleon mass $m_N$ and the box size $L$ is multiplied by the pion mass $M_{\pi}$. The red
solid lines denote the numerical results of $E$ for the respective energy levels while the blue dashed lines denote the
free energy levels of the pion-nucleon final states, i.e. 
\begin{align}
E_{\pi N}^{\rm free} \left( \vec{n} \right) = \sqrt{m_N^2 + \left( \frac{2 \pi}{L} \right)^2 | \vec{n} |^2 } + \sqrt{M_{\pi}^2
+ \left( \frac{2 \pi}{L} \right)^2 | \vec{n} |^2 } \; , 
\end{align}
for $|\vec{n}|^2 = 1,2,3,4$. We can clearly see signs of avoided level crossing at small box sizes, whereas the curves
asymptotally approach the free energy levels at larger box sizes. Also the curves seem to switch between different free
energy levels which is also a typical behaviour for a resonance (see Ref.~\cite{Bernard:2007cm}). It can be especially
observed in the upmost curve between the $|\vec{n}|^2 = 3$ and $|\vec{n}|^2 = 4$ levels. This is exactly the energy
region where the Roper resonance is found, i.e. $\SI{1365}{\MeV}/m_N \approx 1.45$ (which is called the ``critical value'' from here
on) and the curves approximate more and more the free energy levels at energies below the critical value.

\begin{figure}[t]
	\begin{center}  
		\includegraphics[width=0.60\linewidth]{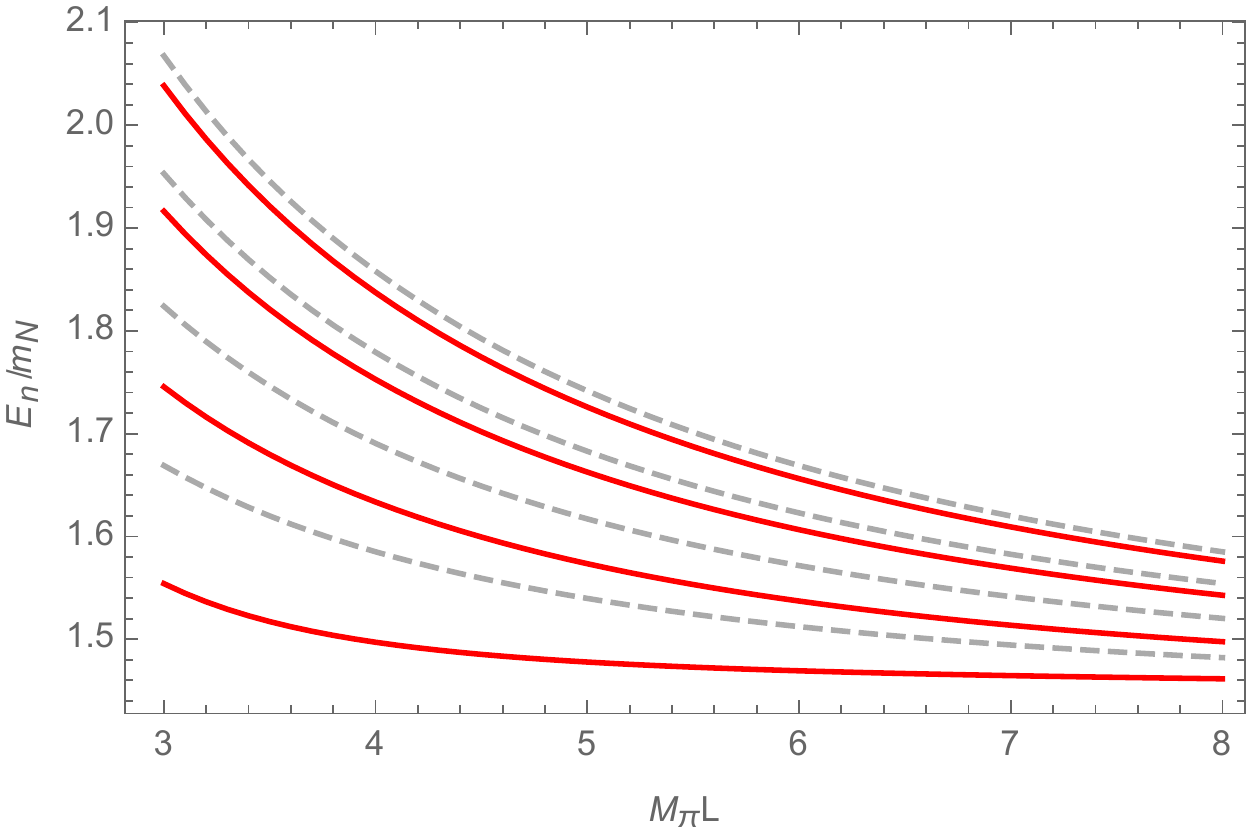}
	\end{center}
	\caption{
	  Energy levels for different box sizes $L$ considering only pion and delta baryon as intermediate states. Red
          solid lines display the numerical results and grey dashed lines the free energy levels of the pion and delta for
          $|\vec{n}|^2 = 1,2,3,4$ (lowest to highest curve). 
	} 
	\label{fig:FitonlyDel}
\end{figure}

Now we will do the opposite and set $g_{\pi N R}=0$. The calculation is performed like before and is displayed in Fig.~\ref{fig:FitonlyDel}. 
\noindent
The free energy levels of the pion and delta in the final state, i.e. 
\begin{align}
E_{\pi \Delta}^{\rm free} \left( \vec{n} \right) = \sqrt{m_{\Delta}^2 + \left( \frac{2 \pi}{L} \right)^2 | \vec{n} |^2 } + \sqrt{M_{\pi}^2
+ \left( \frac{2 \pi}{L} \right)^2 | \vec{n} |^2 } \; ,  
\end{align}
are denoted by the grey dashed lines. This time we see no clear evidence for an avoided level crossing. One reason for this is
the fact that we are now in an energy region which is mostly above the critical value. Only the two lowest lying energy levels
come close to this energy. Another reason is the relatively large coupling $h_R$, which tends to ``wash out'' the typical
signature of avoided level crossing. This effect has been also observed in the energy levels of the delta resonance
in a box~\cite{Bernard:2007cm}.
It is important to note that although the delta baryon is a resonance itself we treat it here as a stable particle.
This holds as a first approximation with the argument that the Roper first decays in a pion and a delta baryon
(or pion and nucleon) and then later
the delta can decay further. For future investigations we should take the unstable nature of the delta into account. One example
to achieve this can be the replacement of the delta propagator in our calculations with a modified propagator including the decay
width of the delta. This will be done in a forthcoming work.

\begin{figure}[t]
	\begin{center}  
		\includegraphics[width=0.60\linewidth]{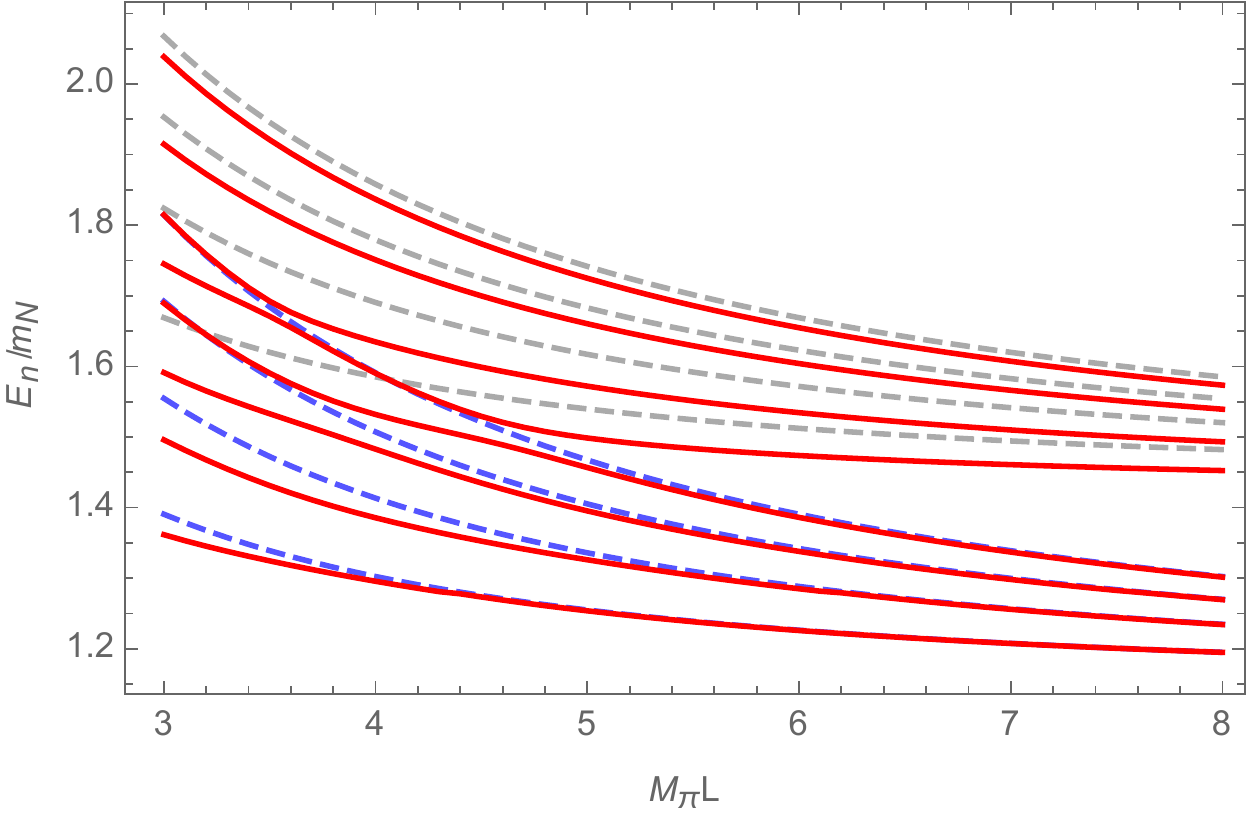}
	\end{center}
	\caption{
	  Energy levels for the full system for different box sizes $L$. Red solid lines display the numerical results and
          blue dashed lines, grey dashed lines display the free energy levels of the pion and nucleon, pion and delta, respectively.  
	} 
	\label{fig:FitFull1}
\end{figure}

\begin{figure}[t]
	\begin{center}  
		\includegraphics[width=0.60\linewidth]{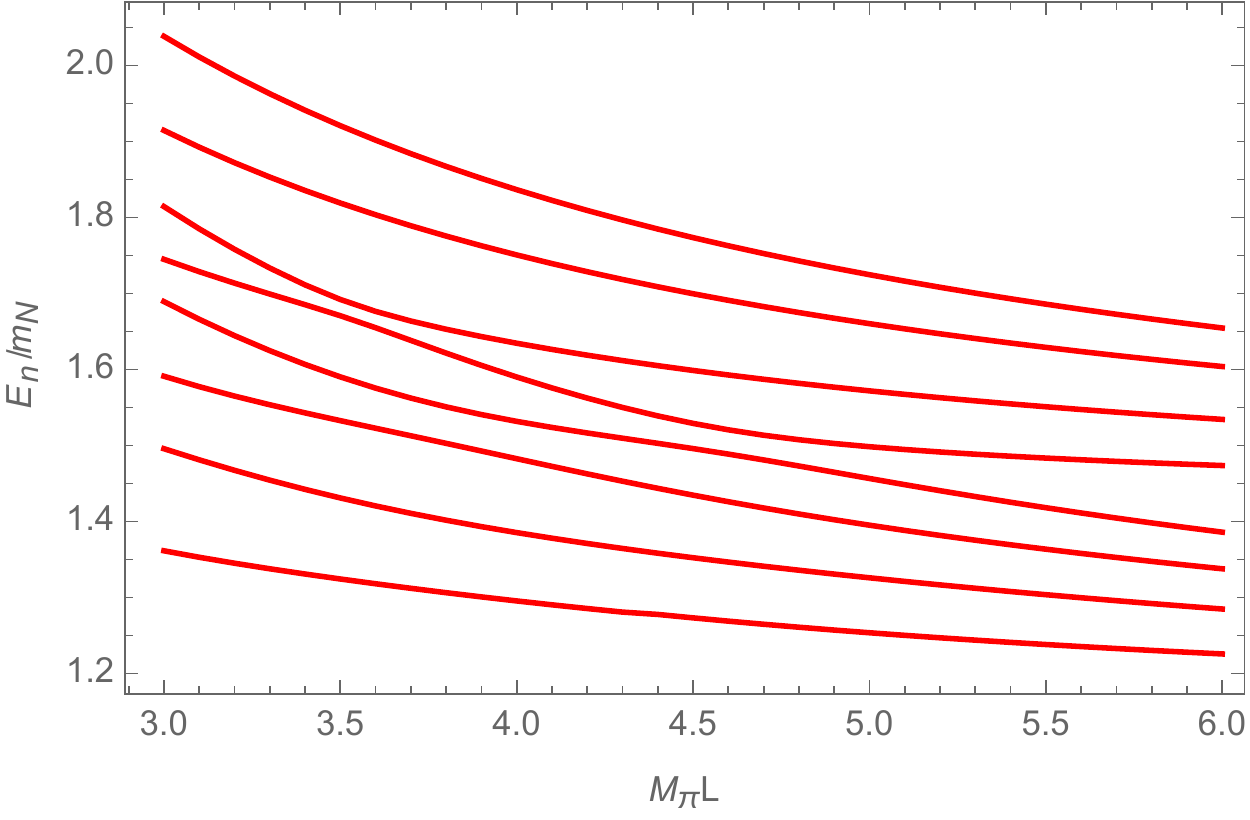}
	\end{center}
	\caption{
		Energy levels for the full system for different box sizes $L$ without displaying the free energy levels. 
	} 
	\label{fig:FitFull2}
\end{figure}

Now we take a look at the full system with pions, nucleons and deltas. Our results are given in Fig.~\ref{fig:FitFull1},
which now include both possible interactions. 
The avoided level crossing is again visible at small box sizes and most pronounced between the free $|\vec{n}|^2 = 3$
level of the pion and the nucleon and the free $|\vec{n}|^2 = 2$ level  of the pion and the delta. The switching of
the energy levels between different free energy levels is clearly seen in the vicinity of the critical value. We also
depict the  results without the free energy levels in Fig.~\ref{fig:FitFull2} to better display the shape of the
curves in this energy region. 
Further away from the critical value and for larger box sizes the energy levels behave like the free ones. Looking at the part of
the fit with small box sizes, one may ask the question what happens at $M_{\pi} L$ values smaller than the ones depicted.
Going to smaller box sizes is problematic because of two things. One is the fact that for small box sizes around
$M_{\pi} L \simeq 3$ the numerical calculation is already quite unstable due to the overlapping energy levels. At smaller
$M_{\pi} L$ it will be extremely difficult to distinguish between the different levels. The other reason is that at smaller
box sizes the exponentially suppressed contributions from the tadpoles and the $\pi R$ loop can not be neglected any more and
have to be considered explicitely.

All in all the energy levels behave according to our expectations and we see the typical signature of avoided level crossing
due to a resonance. A next possible step would be the investigation of the energy levels with the inclusion of an
unstable delta resonance propagator. Also a calculation beyond chiral order $\mathcal{O} \left( p^3 \right)$ should be considered.

\section{Summary and conclusions}
\label{sec:out}
In this paper, we have analyzed the Roper resonance in a finite volume. The calculation of the Roper self-energy up
to third chiral order in the infinite volume has been repeated and the extension to the finite volume case has been achieved
to find the finite volume corrections of the system. We have seen that the FV correction of the self-energy contains
exponentially suppressed contributions for large $L$, which we neglected, and contributions with poles that have to be
regularized. The  calculation of the energy levels has been performed using physical baryon and pion masses and only two LECs had
to be taken into account, which had been determined earlier~\cite{Gegelia:2016xcw}. The main results are: 
\begin{itemize}
\vspace{-2mm}  
\item In the delta-free case ($h_R = 0$) the avoided level crossing can be clearly seen in the vicinity of the Roper
resonance energy. For large box sizes, the energy levels approach the free energy levels. 
\vspace{-2mm} 
\item In the nucleon-free case ($g_{\pi N R} = 0$) there are no clear signs for avoided level crossing. This is caused by the
large value of $h_R$ and by the fact that the energy region lies mostly above the Roper. The approach to the free energy levels
for large $L$ is not as explicit as in the delta free case. 
\vspace{-2mm} 
\item Looking at the full system with nucleons and deltas, the avoided level crossing is observed again.
Also in this case the approach to the free energy levels for large $L$ can be seen.
\end{itemize}
Note that all the discussed  calculations here can also be performed at non-physical pion masses. 
The remaining question is the treatment of the delta resonance in the finite volume. Assuming the delta to be a
stable particle is a reasonable first approximation, but in further calculations its resonance characteristic must be
included. Further, a calculation to fourth order (or higher) in the chiral expansion can be considered. However,
this will increase the number of LECs that have to be taken into account.

\section*{Acknowledgemets}

We are grateful to Akaki Rusetsky for many useful discussions and Jambul Gegelia
for colla\-bo\-ra\-tion in the initial stage of this project and a careful reading of the manuscript.
This work was supported in part by DFG and NSFC through funds provided to the
Sino-German CRC 110 ``Symmetries and the Emergence of Structure in QCD" (NSFC
Grant No.~11621131001, DFG Grant No.~TRR110),  by VolkswagenStiftung (Grant no. 93562)
and by the CAS President's International Fellowship Initiative (PIFI) (Grant No.~2018DM0034).


\appendix
\section{Passarino-Veltman Integrals} \label{AppPVIntegrals}
The Passarino-Veltman Integrals  \cite{Passarino:1978jh} (see also Ref.~\cite{Ellis:2011cr})
are a specific representation of loop integrals,
which we use here. The infinities emerging from the evaluation of the loop integrals in dimensional regularization are contained
in $R$, which is given by 
\begin{align*}
	R = \frac{2}{D-4} - \left[ \log (4 \pi) + \Gamma' (1) + 1 \right] \; ,
\end{align*}
where $D$ denotes the space-time dimension and $\Gamma$ is the Gamma function. This term will be cancelled in the
$\widetilde{\text{MS}}$ renormalization scheme  which is commonly used in ChPT calculations.

The following list contains the loop functions that appear in our calculations and gives their respective results
in the infinite volume.
\begin{itemize}
	\item Integral with one propagator:
	\begin{align*}
	\begin{split}
	A_0 (m^2) &= -16\pi^2 i \mu^{4-D} \int \frac{d^D k}{(2 \pi)^{D}} \frac{1}{k^2 - m^2 + i \epsilon} \\
	&= -m^2  \left[ R +  \log \left( \frac{m^2}{\mu^2} \right) \right] \; .
	\end{split}
	\end{align*}
	
	\item Integral with two propagators:
	\begin{align*}
	\begin{split}
	B_0 (p^2, m^2, M^2) &= -16\pi^2 i \mu^{4-D} \int \frac{d^D k}{(2 \pi)^{D}} \frac{1}{[k^2 - m^2 + i \epsilon] [(k-p)^2 - M^2 + i \epsilon] } \\
	&= (-1) \left[ R - 1 + \log \left( \frac{m^2}{\mu^2} \right) + \frac{p^2 - m^2 + M^2}{p^2} \log \left( \frac{M}{m} \right) \right.  \\
	& \phantom{= (-1) =} \left. + \frac{2 m M}{p^2} F(\Omega) \right] \; ,
	\end{split}
	\end{align*}
	where 
	\begin{align*}
	F(\Omega) = \left\lbrace
	\begin{array}{ll}
	\sqrt{\Omega^2-1} \log ( - \Omega - \sqrt{\Omega^2-1} ) \; , & \Omega \le -1 \\
	\sqrt{1- \Omega^2} \arccos(-\Omega) \; , & -1 \le \Omega \le 1 \\
	\sqrt{\Omega^2-1} \log ( \Omega + \sqrt{\Omega^2-1} ) - i \pi \sqrt{\Omega^2-1} \; , & 1 \le \Omega
	\end{array} 
	\right. \; ,
	\end{align*}
	and
	\begin{align*}
	\Omega = \frac{p^2 - m^2 - M^2}{2 m M} \; .
	\end{align*}
	
	\item Tensor integrals with two propagators: 
	\begin{align*}
	\begin{split}
	B^{\mu} (p^2, m^2, M^2) &= -16\pi^2 i \mu^{4-D} \int \frac{d^D k}{(2 \pi)^{D}} \frac{k^{\mu}}{[k^2 - m^2 + i \epsilon] [(k-p)^2 - M^2 + i \epsilon] } \\
	&:= p^{\mu} B_1 (p^2, m^2, M^2) \; , 
	\end{split}
	\end{align*}
	where
	\begin{align*}
	B_1 (p^2, m^2, M^2) = \frac{1}{2 p^2} \Big\{ \left[ p^2 + m^2 - M^2 \right] B_0 (p^2, m^2, M^2) - A_0 (m^2) + A_0 (M^2)  \Big\} \; , 
	\end{align*}
	and
	\begin{align*}
	\begin{split}
	B^{\mu \nu} (p^2, m^2, M^2) &= -16\pi^2 i \mu^{4-D} \int \frac{d^D k}{(2 \pi)^{D}} \frac{k^{\mu} k^{\nu} }{[k^2 - m^2 + i \epsilon] [(k-p)^2 - M^2 + i \epsilon] } \\
	&:= g^{\mu \nu} B_{00} (p^2, m^2, M^2) + p^{\mu} p^{\nu} B_{11} (p^2, m^2, M^2) \; 
	\end{split}
	\end{align*}
	with 
	\begin{align*}
		B_{00} (p^2, m^2, M^2) &= \frac{1}{2 (D-1)} \Big\{ 2 m^2 B_0 (p^2, m^2, M^2) + A_0 (M^2) \\
		& \phantom{ B_0 (p^2, m^2, M^2)} - \left[ p^2 + m^2 - M^2 \right] B_1 (p^2, m^2, M^2) \Big\} \; , 
	\end{align*}
	and
	\begin{align*}
	B_{11} (p^2, m^2, M^2) &= \frac{1}{2 p^2} \Big\{  \left[ p^2 + m^2 - M^2 \right] B_1 (p^2, m^2, M^2) \\ 
	& \phantom{ B_0 (p^2, m^2)} + A_0 (M^2) - 2 B_{00} (p^2, m^2, M^2)  \Big\} \; .
	\end{align*}
	
	\item Integral with three propagators: 
	\begin{align*}
	\begin{split}
	C_0 (0, p^2, p^2, m^2, m^2, M^2) &= i \mu^{4-D} \int \frac{d^D k}{(2 \pi)^{D}} \frac{-16\pi^2}{ \left[ k^2 - m^2 + i \epsilon \right]^2  \left[ (k-p)^2 - M^2 + i \epsilon \right]^{} } \\
	&= \left(\frac{1}{2 m}\right) \frac{\partial}{\partial m} B_0 (p^2, m^2, M^2) \; ,
	\end{split}
	\end{align*}
	and 
	\begin{align*}
	\begin{split}
	C_1 (0, p^2, p^2, m^2, m^2, M^2) &= \frac{1}{4 p^2} \left[ B_0 (p^2, m^2, M^2) - B_0 (0, m^2, m^2) \right. \\ 
	& \phantom{= \frac{1}{4 p^2} = } \left. + (m^2 - p^2 - M^2) C_0 (0, p^2, p^2, m^2, m^2, M^2) \right] \; .
	\end{split}
	\end{align*}
\end{itemize}
There are also some special cases appearing, which are listed below
\begin{align*}
B_0 (0, m^2, m^2) = (-1) \left[ R +1 + \log \left( \frac{m^2}{\mu^2} \right) \right] \; ,
\end{align*}
\begin{align*}
B_0 (m^2,0 , m^2) = (-1) \left[ R -1 + \log \left( \frac{m^2}{\mu^2} \right) \right] \; .
\end{align*}
These relations can be shown with the explicit form of $B_0$ and it follows that 
\begin{align*}
B_0 (0, m^2, m^2) +2 =  B_0 (m^2,0 , m^2) \; .
\end{align*}

\section{Useful Formulas} \label{AppUsefformula}
This section contains a handful of useful formulas that were used in our calculations. 
\begin{itemize}
	\item Feynman parameter: 
		\begin{align*}
			\frac{1}{AB} = \int_{0}^{1} \frac{dy}{\lbrack yA + (1-y) B \rbrack^2} \; . 
		\end{align*}
	\item Modified Bessel functions of the second kind (see, e.g., \cite{Abram1}): 
		\begin{align*}
			K_{\nu} (z) := \int_{0}^{\infty} dt \cosh \left( \nu t \right) e^{ - z \cosh (t)} \; , \quad \text{for} \; z>0  \; . 
		\end{align*}
		Special case:   
		\begin{align*}
			K_0 (z) = \int_{0}^{\infty} dt \frac{\cos (z t)}{\sqrt{t^2 +1}} \; , \quad \text{for} \; z>0  \; . 
		\end{align*}
\end{itemize}

\end{document}